\documentclass{ws-p9-75x6-50}
\input epsf.tex
\def\DESepsf(#1 width #2){\epsfxsize=#2 \epsfbox{#1}}

\begin{document}

\title{CP Violaing Phases In SUSY GUT Models\footnote{Based on talk at
SSS-99, Seoul, Korea, June, 1999} } 
\author{ E. Accomando, R. Arnowitt and B. Dutta }
\address{ Center For Theoretical Physics, Department of Physics, Texas A$\&$M
University, College Station TX 77843-4242}
\date{January, 2000}
\maketitle
\abstracts {Supersymmetric CP violating phases are examined within the
framework of gravity mediated supergravity grand unified models with R parity
invariance for models with a light ($\stackrel{<}{\sim} 1$ TeV) particle
spectrum. In the minimal model, the nearness of the t quark Landau pole
naturally suppresses the t-quark cubic soft breaking parameter at the
electroweak scale allowing the electron and neutron experimental electric
dipole moment (EDM) constraints to be satisfied with a large GUT scale phase.
However, the EDM constraints require that $\theta_B$, the quadratic soft
breaking parameter phase be small at the electroweak scale unless
tan$\beta\stackrel{<}{\sim}3$, which then implies that at the GUT scale this
phase must be large and highly fine tuned to satisfy radiative breaking of
$SU(2)\times U(1)$. Similar results hold for non minimal models, and a possible
GUT model is discussed where all GUT scale CP violating phases are naturally
small (i.e. O($10^{-2})$). An interesting D-brane model is examined which
enhances the size of the phases over much of the parameter space at the
electroweak sector for tan$\beta\stackrel{<}{\sim} 5$, but still possesses the
fine tuning problem at the GUT scale.}

\section {Introduction}
The Standard Model (SM) of strong and electroweak
interactions is a remarkably rigid theory of quark and lepton interactions. Thus
the gauge invariance guarantees baryon and lepton number invariance and forbids
bare mass terms. In order to generate quark and lepton masses, one introduces
Higgs Yukawa couplings, and the spontaneous breaking of $SU(2)\times U(1)$
simultaneously generates the necessary masses, both in the gauge boson and
fermionic sectors. For the three generations that have been observed, the Yukawa
matrices allow for only one CP violating phase in the CKM matrix, which is
consistent with the observed CP violation in the K meson system (and these ideas
will be tested for B mesons with future data from B factories and high energy
accelerators.). Further this phase gives only a very small contribution to the
electron and neutron electric dipole moments (EDMs), $d_e$ and $d_n$. In
supersymmetry (SUSY) extensions of the SM, things are more complicated. One
generally imposes R-parity invariance to suppress too rapid proton decay, and
there is now an array of possible new CP violating phases arising from the SUSY
soft breaking masses which generally produce large contributions to $d_n$ and
$d_e$. The current experimental EDM bounds are for $d_n$\cite{dn} and $d_e$
\cite{de}: \begin{eqnarray}
(d_n)_{exp}&<&6.3\times 10^{-26}ecm;
\,90\% C.L.\\
(d_e)_{exp}&<&4.3\times 10^{-27}ecm; \,95\% C.L.\label{dnde}\end{eqnarray}

In the past, two suggestions have been made to accomodate these bounds: (i) one
can assume the CP violating phases are small i.e. $\phi=O(10^{-2})$\cite{redm3} and/or (ii) the SUSY mass spectrum is heavy i.e. squark ($\tilde q$), slepton
($\tilde l$) and gluino ($\tilde g$) masses are $\stackrel{>}{\sim}O(1 {\rm
TeV})$
\cite{redm4}. The first hypothesis would appear to require a significant amount of
fine tuning(though see sec. 4 below), while the second would place the SUSY
spectrum beyond the reach of even the LHC and regenerate the gauge hierarchy
problem. Recently, it has been observed that a third option is possible, i.e.
that cancellations can naturally occur in the total EDM amplitude (e.g. between
neutralino and chargino contributions) allowing one to satisfy Eqs. (1,2) with
relatively large phases (i.e. O($10^{-1}$)) and a light SUSY spectrum
$<O(1{\rm TeV})$ \cite{nath1}., and there has been considerable investigation of this
possibility \cite{nath1,nath2,nath3,falk1,bk,falk2,kane1,bartl,pokorski}. 
The work presented here is given in Ref\cite{aad1,aad2}.

We consider here these possibilities within the framework of supergravity
(SUGRA) grand unified models (GUTs) with gravity mediated SUSY breaking and
R-parity invariance \cite{acn}. For other work within this framework see 
\cite{nath1,nath2,falk1,bk,kane2,bartl}. Here the
low energy predictions of the model are determined from the GUT scale $M_G$
parameters by running the renormalization group equation from $M_G$ to the
electroweak scale $M_{\rm EW}$. Such a theory is considerably more constrained
than the purely phenomenological low energy MSSM model. Thus (1) there are
considerable constraints arising from the GUT group symmetry, (2) CP violating
phases and SUSY parameters that are arbitrary in the MSSM get correlated by the
RGE, (3) radiative breaking of $SU(2)\times U(1)$ at M$_{\rm EW}$ puts additional
constraints on the CP violating phases and SUSY parameters.

In models of this type,  what is natural or unnatural is a property of the
theory at $M_G$ rather than $M_{\rm EW}$. Thus we find that some phases that
are naturally large at $M_G$ get suppressed at $M_{\rm EW}$ leading, hence, to
a ``naturally" small phase there. However, unless tan$\beta\stackrel{<}{\sim}$3
(assuming SUSY masses are not large) one phase at $M_{\rm EW}$ is quite small,
and the RGE then implies that it is both large and highly fine tuned at $M_G$.
In addition, while there is considerable theretical uncertainty in the
calculation of $d_n$, the combined constraints of $d_n$ and $d_e$ of Eqs. (1,2) put significant additional
constraints on the allowed SUSY parameter space. 

The above results hold for both for the minimal mSUGRA model and generally for
models with nonuniversal soft breaking. We will also discuss one very
intersting D-brane model with nonuniversal soft breaking \cite{kane2}.

\section{Electric Dipole Moments For mSUGRA Models}
We consider first the simplest case where there is universal SUSY soft breaking
parameters at $M_G$. The theory then depends on five parameters: $m_0$ (the
universal squark and slepton mass at $M_G$), $m_{1/2}$ (the universal gaugino
mass at $M_G$), $A_0$ (the cubic soft breaking parameter at $M_G$), B$_0$(the
quadratic soft breaking parameter at $M_G$) and $\mu_0$ (the Higgs mixing
parameter at $M_G$ in the superpotential). Of these the last four can be
complex. However, one may make a phase rotation to make $m_{1/2}$ real, leaving $A_0$, $B_0$ and $\mu_0$ complex at $M_G$:
\begin{eqnarray} A_0&=&|A_0|e^{i \alpha_{0A}};\, B_0=|B_0|e^{i
\theta_{0B}};\,\mu_0=|\mu_0|e^{i
\theta_{0\mu}}.
\label{amb}\end{eqnarray}
The RGE  determines the low energy parameters in terms
of the $M_G$ parameters. Thus there results a different A parameter at
$M_{\rm EW}$ (which we take to be $m_t$), for each squark and slepton e.g. 
$A_t$, $A_b$,$A_\tau$,$A_u$,
$A_d$,
$A_e$ and we label these by 
\begin{eqnarray}A_t&=&|A_t|e^{i \alpha_{t}};\, B=|B|e^{i
\theta_{B}};\,\mu=|\mu|e^{i
\theta_{\mu}}; etc
\label{amb1}\end{eqnarray}
(Note that to one loop order $\theta_{0\mu}=\theta_{\mu}$.) The $SU(3)_C$,
$SU(2)_L$ and $U(1)_Y$ gaugino masses are labeled 
$\tilde m_3$, $\tilde m_2$ and $\tilde m_1$ and remain real at one loop..

\begin{figure}[htb]
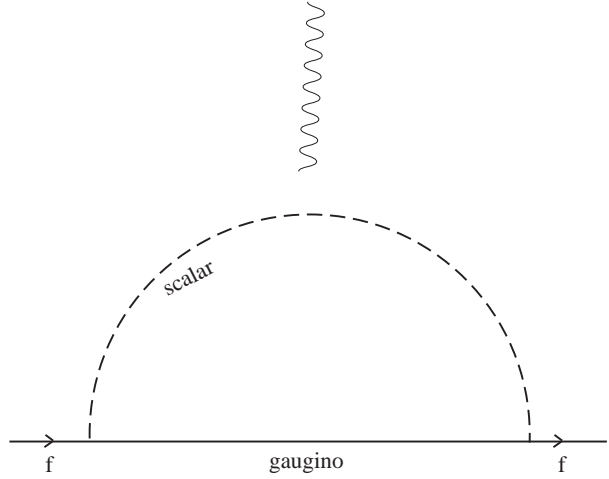

\centerline{ \DESepsf(edmfig1.epsf width 8 cm) }
\smallskip
\caption {\label{fig1} Diagrams contributing to $d_f$. For f=q(l) the scalars
are $\tilde q(\tilde l)$ and gauginos are $\tilde\chi^0_i$, $\tilde\chi^\pm_i$,
$\tilde g$ where $\tilde\chi^\pm_i$, i=1,2 are charginos and
$\tilde\chi^0_i$, i=1...4 are neutralinos.  }
\end{figure}

The effective Lagrangian for $d_f$, the EDM of  fermion of type f (quark or
lepton) is 
\begin{eqnarray}  L_f&=&-{i\over 2}d_f\bar f\sigma_{\mu\nu}\gamma^5 f F^{\mu\nu}.
\label{lag}
\end{eqnarray} The basic diagrams for  $d_f$ that contribute to $L_f$ at
$M_{\rm EW}$ are given  in Fig.\ref{fig1}. 

The calculation of the neutron EDM contains a number of  uncertainties due to 
QCD effects. To relate $d_n$ to the quark EDMs $d_u$ and $d_d$, we use 
the non relativistic quark model relation
\begin{eqnarray}  d_n&=&{1\over 3}(4d_d-d_u)
\label{qm}
\end{eqnarray}. In addition to Eq.\ref{lag},
one must take into account the gluonic operators $L^C$ and $L^G$ given by
\begin{eqnarray}L^C&=&-{i\over 2}d^C\bar
q\sigma_{\mu\nu}\gamma^5T^aqG_a^{\mu\nu}
\label{lc}
\end{eqnarray} \begin{eqnarray}  L^G&=&-{1\over
3}d^Gf_{abc}G_{a\mu\alpha}G_{b\nu}^{\alpha}\tilde{G}_c^{\mu\nu}
\label{lg}
\end{eqnarray}  where $\tilde{G}^{\mu\nu}_c= {1\over
2}\epsilon^{\mu\nu\alpha\beta}G_{c\alpha\beta}$ $(\epsilon^{0123}=1)$,
$T^a={1\over 2}\lambda^a$(
$\lambda^a$ are the SU(3) Gell-Mann matrices), ${G}^{\mu\nu}_a$ are
the SU(3) field strengths and 
$f_{abc}$ are the SU(3) structure constants. Contributions to $L^C$ arise from
 Fig.\ref{fig1} with $\gamma$ replaced by $g$ and the two loop
Barr-Zee type diagrams of Fig.\ref{fig2}\cite{ckp}. Contributions to $L^G$ come
from the  two loop Weinberg type diagram of Fig.\ref{fig3} \cite{dai}.
One must use the QCD RGE factors
$\eta^{ED}$, $\eta^G$, $\eta^C$ to evolve the results from $M_{\rm EW}$
down to 1 GeV \cite{aln} and we use the naive dimensional analysis\cite{mon} to
relate $d^C$ and $d^G$ to $d_f$. In calculating $d_q$, one needs to know
the quark masses $m_u$ and $m_d$. While the mass ratios are fairly well known
\cite{lr},
\begin{eqnarray}  {m_u\over m_d}&=&0.553\pm 0.043;\,\,\,\, {m_s\over
m_d}=18.9\pm 0.8
\label{qr}
\end{eqnarray} $m_s$ is in considerable doubt, i.e. QCD sum rules give 
$m_s=(175\pm 25)$ MeV and lattice gauge theory gives 
$m_s(2 GeV)=(100\pm 20 \pm 10)$ MeV  in the quenched lattice
calculation. (Lowering the scale to 1 GeV will increase $m_s$, but
unquenching will reduce it).

We see in general that there are significant uncertainties in calculating $d_n$
( perhaps a factor of 2-3). In the following we will assume $m_s=150$ GeV
($m_d\cong$ 8 MeV, $m_u\cong$ 4.4 MeV), but we will exhibit below the sensitivity
of $d_n$ to the uncertainty in $m_s$.

$SU(2)\times U(1)$ breaking at $M_{\rm EW}$ gives rise to Higgs VEVs which
in general may be complex:
\begin{eqnarray}<H_{1,2}>=v_{1,2}e^{i\epsilon_{1,2}};\,
v_{1,2}\equiv|<H_{1,2}>|\end{eqnarray} and we define $tan\beta=v_2/v_1$.
One may chose matter phases so that the chargino
mass matrix takes the form:
\begin{eqnarray} M_{\chi^{\pm}}&=&\left(\matrix{
 \tilde m_2                & \sqrt 2 M_W sin\beta  \cr
  \sqrt 2 M_W cos\beta            &-|\mu|e^{i\theta} }\right)\label{char}
\end{eqnarray}
where $\theta=\epsilon_1+\epsilon_2+\theta_\mu$. The neutralino mass matrix is:
\begin{eqnarray} M_{\chi^0}&=&\left(\matrix{
 \tilde m_1  &0            &a   &b\cr
  0          &\tilde m_2   &c   &d  \cr
  a          &c            &0   &|\mu|e^{i\theta}\cr
  b          &d            &|\mu|e^{i\theta}   &0  \cr}\right)\label{neut}
\end{eqnarray}
  where $a=-M_Z sin\theta_W cos\beta$, $b=M_Z sin\theta_W sin\beta$,
$c=-cot\theta_W a$, $d=-cot\theta_W b$. The squark mass matrices is
\begin{eqnarray} M_{\tilde q}^2&=&\left(\matrix{
 m^2_{q_L}               & e^{-i\alpha_q}m_q(|A_q|+|\mu| R_q
e^{i(\theta+\alpha_q)}) 
\cr
  e^{i\alpha_q}m_q(|A_q|+|\mu| R_q e^{-i(\theta+\alpha_q)})            
&m^2_{q_R} }\right)\label{sqrk}
\end{eqnarray} where $m_q$, $e_q$ are the quark mass and 
charge,\begin{eqnarray} m_{q_L}^2&=&m^2_Q+m_q^2+(1/2-e_q sin^2\theta_W)M_Z^2
cos2\beta\\ m_{q_R}^2&=&m^2_U+m_q^2+e_q sin^2\theta_W M_Z^2 cos2\beta\label{qlqr}
\end{eqnarray} where $R_q=cot\beta(tan\beta)$ for u(d) quarks and 
$m_{Q}^2$, $m_{U}^2$ are given in
\cite{iba}. (A Similar result
holds for the sleptons).

The Higgs VEVs are determined by minimizing the
effective potential  
\begin{eqnarray}
V_{eff}&=&m_{1}^2v_1^2+m_{2}^2v_2^2+2|B\mu|cos(\theta+\theta_B)v_1v_2+
{g^2_2\over 8}(v_1^2-v_2^2)^2\\\nonumber&+&{{g^{\prime}}^2\over
8}(v_2^2-v_1^2)^2+V_1\label{veff}
\end{eqnarray} where $V_1$ is the one loop contribution 
\begin{eqnarray}V_1={1\over {64 \pi^2}}\sum_a C_a(-1)^{2 j_a} (2 j_a+1) m_a^4
(ln{m_a^2\over Q^2}-{3\over 2}).\label{v1}
\end{eqnarray} Here $C_a$, $j_a$ and $m_a$ are the color factor, spin and mass of
particle $a$ and $Q=m_t$ is the electroweak scale. In the following we include
the full third generation of states in $V_1$ (t, b $\tau$) so that we can
consider large tan$\beta$. The minimization of the tree contribution gives 
\begin{eqnarray}\theta=\pi-\theta_B.\label{theta}
\end{eqnarray} From Eq.(\ref{sqrk}), the mass eigenvalues entering into
Eq.(\ref{v1}) depends only on 
$\theta+\alpha_q$ and
$\theta+\alpha_l$, so that minimizing the tree plus loop $V_{\rm eff}$ gives  
\begin{eqnarray}
\theta=\pi-\theta_B+ f_1(\pi-\theta_B+\alpha_q,\pi-\theta_B+\alpha_l
)\label{thetaf}
\end{eqnarray}  where $f_1$ is the one loop correction. As we will see, this
correction can become significant for large tan$\beta$. However, as we will see
below, it can make important contributions for large tan$\beta$ since the EDMs
are sensitive to $\theta_B$.

A convenient way of characterizing how close the theory is in accord with the
EDMs is given by the parameter K:
\begin{eqnarray}  K=log_{10}\mid{d_f\over {(d_f)_{exp}}}\mid
\label{k}
\end{eqnarray} where $(d_f)_{exp}$ is the current experimental bound. Thus
$K\leq 0$ is required for the theory to be in accord with experiment. An
example of K as a function of $m_0$ is given in Fig.\ref{fig4}. One sees that
the width of the allowed $K\leq 0$ region decreases as tan$\beta$ increases.
Eventually, for very large $m_0$, one would obtain $K<0$ even for large
tan$\beta$ (the heavy SUSY spectrum option). For $m_0<1$ TeV, we see that the
region $K\leq 0$ moves to ower $m_0$. The reason for this is that the chargino
diagram increases more rapidly with tan$\beta$ than the neutralino diagram, and
in order to maintain the cancelation between them to satisfy $K\leq 0$ one must
decrease $m_0$ to enhance the neutralino diagram relative to the chargino. The
above discussion shows that the cancelations needed to achieve $K\leq 0$  are
relatively delicate, and would become increasingly so if e.g. the experimental
bounds are reduced by a factor of 10 ($K\leq -1$).

To understand what parameters control cancelations, one may look at the RGE
which relate the GUT scale parameters  to
electroweak scale parameters. The one loop RGE must be solved numerically.
However, for small or intermediate tan$\beta$ (and also in the SO(10) limit)
analytic solutions are availbale which allow one to see analytically what is
happenning. Thus for low tan$\beta$ one finds for $A_t$ the result
\begin{eqnarray}A_t&=&D_0A_0+\Phi_A m_{1/2}\label{at}
\end{eqnarray} where $\Phi_A$ is real and O(1) and 
$D_0\cong 1-(m_t/{200 sin\beta})^2$. Thus $D_0$ is  small i.e.
$D_0\stackrel{<}{\sim} 0.2$ . The imaginary part of
Eq.(\ref{at}) gives:\begin{eqnarray}
|A_t|sin\alpha_t&=&|A_0|D_0sin\alpha_{0A}
\label{At2}
\end{eqnarray}
Thus because $D_0$ is small (the nearness of the t-quark Landau pole) the phase
$\alpha_t$ is suppressed relative to $\alpha_{0A}$. Thus even if
$\alpha_{0A}=\pi/2$, $\alpha_t$ is sufficiently reduced at the electroweak
scale so that the EDM constraints may be satisfied. Hence in mSUGRA, no fine
tuning of $\alpha_{0A}$ is necessary.

The situation, however, is more difficult for the $\theta_B$ phase. For low
and intermediate tan$\beta$ (and similar results hold for large tan$\beta$ in
the SO(10) limit) the RGE solution is 
\begin{eqnarray}B&=&B_0-{1\over 2}(1-D_0)A_0+\Phi_B m_{1/2}\label{b}
\end{eqnarray} where $\Phi_B$=real and $O(1)$. The imaginary and real parts give
\begin{eqnarray}|B|sin\theta_B&=&|B_0|sin\theta_{0B}-{1\over
2}(1-D_0)|A_0|sin\alpha_{0A}\label{bsin}
\end{eqnarray}
\begin{eqnarray}|B|cos\theta_B&=&|B_0|cos\theta_{0B}-{1\over
2}(1-D_0)|A_0|cos\alpha_{0A}+\Phi_B m_{1/2}.\label{bcos}
\end{eqnarray} These may be viewed as equations to determine $|B|$ and
$\theta_B$ in terms of the GUT scale parameters. Alternatively, one may impose
phenomenological constraints at the electroweak scale to see what GUT scale
parameters will satisfy them. Two such constraints are the experimental EDM
bounds, and the requirement of radiative breaking of $SU(2)\times U(1)$ at the
electroweak scale. The latter implies that 
\begin{eqnarray}|B|&=&{1\over 2}sin{2\beta}{m_3^2\over|\mu|}\label{bmag}
\end{eqnarray} where $m_3^2=2|\mu|^2+m_{H_1}^2+m_{H_2}^2+ 1 {\rm loop}$, where
$m_{H_i}^2$ are the Higgs running  masses. 

The EDM constraints, generally require that $\theta_B$ be small i.e.
$\theta_B\stackrel{<}{\sim}0.2$. Since ${1\over 2}sin2\beta\simeq(1/tan\beta)$,
 the r.h.s. of Eq.(\ref{bmag}) is a decreasing function of tan$\beta$, and
hence it determines $|B|$ to be small unless $tan\beta\stackrel{<}{\sim}$3.
Returning to Eq.(\ref{bsin}), one sees that the l.h.s. is thus quite small and
so $\theta_{0B}$ is scaled by $\alpha_{0A}$. Hence if $\alpha_{0A}$ is not fine
tuned to be small, we find that the GUT scale $\theta_{0B}$ must be large i.e.
$O(1)$. (This is indeed confirmed by detailed numerical calculations.)
Consider now the situation where one fixes $A_0$ at $M_G$, and let $\Delta\theta_B$ 
be the allowed range of
$\theta_B$ satisfying the EDM constraints. One has
$\Delta\theta_B\stackrel{<}{\sim}0.2$ and in general smaller. From Eq. (24)
we see the corresponding allowed range at $M_G$ is 
\begin{equation}
\Delta\theta_{0B}\cong{|B|\over {|B_0|}}\Delta\theta_B\ll\Delta\theta_B
\label{DthetaB}
\end{equation} and by $SU(2)\times U(1)$ breaking constraint Eq. (26) one will have 
$\Delta\theta_{0B}<<\Delta\theta_B$ unless tan$\beta\stackrel{<}{\sim}$3. 
Fig.5, where the  value of
$\theta_{0B}$ obeying the EDM constraint is plotted as a function of
$m_{1/2}$ for $|A_0|=300$ GeV and $\alpha_{0A}=\pi/2$  illustrates this effect. One sees that
indeed $\theta_{0B}$ is large, and that even for tan$\beta$=3, the allowed range of 
$\Delta\theta_{0B}$ satisfying the $d_e$ constraint for 
$m_{1/2}\stackrel{<}{\sim}$350 GeV ($m_{\tilde g}\stackrel{<}{\sim}$1 TeV) is
quite small. For tan$\beta$=10, $\Delta\theta_{0B}$ is very small. (Note that the LEP
bound on the Higgs mass already requires tan$\beta\geq$2.) 

We see that the combined
requirements of radiative electroweak breaking and the experimental EDM constraints
lead to a serious fine tuning problem at the GUT scale: $\theta_{0B}$ is large and
must be tightly fine tuned unless tan$\beta$ is close to its minimum value
tan$\beta$=2 (unless SUSY masses are large, e.g. $\stackrel{>}{\sim}$ 1 TeV, and/ or all phases are
small, e.g. $O(10^{-2})$. )

We discussed above the existence  of uncertainties in calculating of
the neutron EDM. We exhibit here the effect of the uncertainty in $m_s$ in
Eq.(9). Fig.6  plots $K= 0$ contours corresponding to $m_d$(1
GeV)=5,8 and 12 MeV for tan$\beta$=3, $|A_0|=300$ GeV and 
$\alpha_{0A}=\pi/2$. These values of $m_d$ correspond to $m_s\cong$ 95, 151
and 227 MeV respectively. (Current lattice calculations favor the lowest
value.) We see that $d_n$ is quite sensitive to $m_d$, the constraint being
significantly less severe for the lower quark masses. In Fig.5 and
subsequent figures, we have chosen the central value of $m_d$= 8 MeV.

We also have mentioned above that the loop corrections in Eq. (19) can be
significant for large tan$\beta$. This is because $f_1$ which grows with
tan$\beta$, represents an effective shift in $\theta_B$, and the EDM
constraints require $\theta_B$ to be small. One can see this effect of $f_1$
in Fig. 7 for tan$\beta$=20, where we have set $\theta_B=0$ so that in this
example $f_1$ is the total ``effective" $\theta_B$ in Eq. (19). For large
$m_{1/2}$, the shape of the curves resemble those of Fig.6. However for 
175 GeV$\stackrel{<}{\sim}m_{1/2}\stackrel{<}{\sim}400$ GeV, the $f_1$
contribution gives rise to cancelation in the EDM amplitudes to
significantly reduce the excluded regions showing that the loop
contributions are important for large tan$\beta$. Note that the effect
persists even for $K=-0.5$.

\section{D-Brane Models}
In SUGRA models, nonuniversal soft breaking can arise if the hidden sector fields in
the Kahler potential which give rise to SUSY breaking do not couple universally to the
physical sector fields. In this case nonuniversal squark, slepton and Higgs masses can
be generated at $M_G$, as well as nonuniversal A parameters. For simple GUT groups,
however, it is difficult to generate more than small nonuniversalties in the gaugino
masses at $M_G$. Models of this type behave qualitatively similar to the mSUGRA model
discussed in Sec.II above i.e. no serious fine tuning is needed for $\alpha_{0A}$, but
$\theta_{0B}$ is generally large and highly fine tuned unless 
tan$\beta\stackrel{<}{\sim}$3.

We consider in this section a class of models based on Type IIB orientifolds where
the existence of
open string sectors  imply the presence of $Dp$-branes,
manifolds of p+1 dimensions in the full D=10 space of which 6 dimensions are
compactified e.g. on a six torus $T^6$. (For a general discussion of this class of
 models see \cite{iba2}). Models of this type can contain 9 branes (the
full 10 dimensional space) plus $5_i$-branes, i=1, 2, 3 (6 dimensional space with 
two compact dimensions) or in the T-dual representation, 3 branes plus 7$_i$ branes, i=1, 2, 3 . Associated with a set of
n coincident branes is a gauge group U(n).

One can clearly embed the Standard model gauge group in a number of ways in such
models. Recently, an interesting model has been proposed  based on
9-branes and 5-branes \cite{kane2}. In this model, $SU(3)_C\times U(1)_Y$ is associated
with one set of 5-branes, i.e. $5_1$, and SU(2)$_L$ is associated with a
second intersecting set 5$_2$. Strings starting on 5$_2$ and ending on 5$_1$
have massless modes carrying the joint quantum numbers of the two branes i.e.
 the SM quark, lepton and Higgs doublets. 
Strings beginning and ending on 5$_1$ have massless modes carrying
$SU(3)_C\times U(1)_Y$ quantum numbers i.e. the SM quark and lepton singlets. We
assume all other possible fields at the compactification scale $M_c$ are superheavy and
can be ignored to  first approximation, as far as the low energy predictions of the
model are concerned. For models of this type $M_c=M_G$, while the string scale
 $M_{\rm
str}$, is given by $M_{\rm str}=(\alpha_G M_c
M_{Planck}/\sqrt{2})^{1/2}\cong8\times 10^{16}$ GeV (for $\alpha_G\cong1/24$). Thus
below $M_G$, the gauge interactions are the ususal D=4 theory.

The gauge kinetic functions for 9 branes and 5$_i$-branes are given by
\cite{iba2,iba1}
$f_9=S$ and $f_{5_i}=T_i$ where S is the dilaton and $T_i$ are moduli. The
origin of SUSY breaking is not yet understood
in string theory. It may be parametrized, however, by VEV growth of $S$ and $T_i$. 
Further, in string theory, CP violation must also occur as a spontaneous
breaking and it is natural to associate thses two spontaneous breakings by assuming
that   the F-components grow complex  VEVs which are parametrized by \cite{iba2,iba3,iba4}
\begin{eqnarray} F^S&=&{2\sqrt{3}}<{\rm Re} S>{\rm
sin}\theta_b e^{i\alpha_s}m_{3/2}\\\nonumber 
F^{T_i}&=&{2\sqrt{3}}<{\rm Re} T_i>{\rm
cos}\theta_b \Theta_i e^{i\alpha_i}m_{3/2}
\label{fsft}
\end{eqnarray}
where $\theta_b$, $\Theta_i$ are Goldstino angles
($\sum\Theta_i^2=1$) and $m_{3/2}$ is the gravitino mass. 
In the following we  assume  $\Theta_3=0$, $<Re T_i>$ are
equal (to guarantee grand unification at $M_G$), and $<Im T_i>$=0 (so that the
spontaneous breaking does not grow a $\theta$-QCD type term).

The above model then leads to the following soft breaking masses at $M_G$:
\begin{eqnarray}  \tilde{m_1}&=&\sqrt{3}{\rm
cos}\theta_b\Theta_1e^{-i\alpha_1}m_{3/2}=\tilde{m_3}=-A_0\\
\tilde{m_2}&=&\sqrt{3}{\rm
cos}\theta_b\Theta_2e^{-i\alpha_2}m_{3/2}
\label{m1m2}
\end{eqnarray}
and
\begin{eqnarray}  m_{5_15_2}^2&=&(1-{3\over 2}{\rm sin}^2\theta_b) m^2_{3/2}\\
 m_{5_1}^2&=&(1-{3}{\rm sin}^2\theta_b) m^2_{3/2}
\label{m512}
\end{eqnarray}
where $m_{5_15_2}^2$ are the
soft breaking masses for $q_L$, $l_L$, $H_{1,2}$ and $m_{5_1}^2$ are for
$u_R$, $d_R$ and $e_R$. The $B_0$ and $\mu_0$ parameters are  not determined by the above considerations and
are model independent. We therefore
parametrize them phenomenologically by
\begin{eqnarray}  B_0=|B_0|e^{i
\theta_{0B}};\,\mu_0=|\mu_0|e^{i
\theta_{0\mu}}.
\label{mb}
\end{eqnarray}
 We can also chose phases such that $\alpha_2=0$. 
 
 We see that the D-brane model give rise to a soft breaking pattern uniquely different
 from what is seen in SUGRA GUT models. Thus it would be difficult to find a GUT group
 breaking where $\tilde m_1=\tilde m_3\neq \tilde m_2$  and and
similarly the above pattern of sfermion and Higgs soft masses. Brane models
can achieve the above pattern since they have the freedom of associating different
parts of the SM gauge group with different branes. In particular, the fact that the $\tilde m_1$ amd $\tilde m_3$ phases are
 equal causes cancelation between the gluino and neutralino EDM diagrams, and
 considerably aids in satisfying the EDM constraints. Fig.8 exihibits this phenomena
 where  K is plotted as a function of $\theta_B$ for $d_e$
for tan$\beta$=2 (solid), 5 (dashed), 10 (dotted) with phases
$\phi_1=\phi_3=\pi+\alpha_{0A}=-1.25\pi$ and $m_{3/2}=150$ GeV, $\theta_b=0.2$,
$\Theta_1$=0.85. One sees that $\theta_B$ can be quite large and still satisfy
the EDM bound $K\le$0, i.e. $\theta_B\simeq0.4$ for tan$\beta$=2 and 
$\theta_B\simeq0.25$ even for tan$\beta$=10. If one reduces the gaugino phases,
e.g. to  $\alpha_1=\alpha_3=-1.1\pi$, one finds $\theta_B<0.2$ showing that the
enhanced values of $\theta_B$ are indeed due to the cancellations allowed by the
gaugino phases.

While $\theta_B$ can be relatively large at the electroweak scale, one finds
as in the mSUGRA models, $\theta_{0B}$ at the GUT scale is large (when
$\alpha_1=\alpha_3=\pi+\alpha_{0A}$ is large) but must be fine tuned, i.e.
the allowed range $\Delta\theta_{0B}$ is small unless tan$\beta$ is small.
This is illustrated in Fig.9 where $\Delta\theta_{0B}$ is plotted as a
function of tan$\beta$. For these parameters ($\alpha_{0A}=-\pi/4$) there is
already a 1 $\%$ finetuning in $\theta_{0B}$ at tan$\beta$=5, with increased
fine tuning required for higher tan$\beta$. Thus this class of D-brane
models does not resolve the fine tuning problem.

While as discussed in Sec.2 there are uncertainties in the calculation of
the neutron EDM, it is of interest to see what parts of the parameter space
remains if one requires that the experimental EDM constraint is satisfied
simultaneously for $d_e$ and $d_n$. (We assume here the validity of the
calculations of $d_n$ described in sec.2) An example of this is shown in
Fig.10 where the allowed regions for $d_e$ and $d_n$ are shown for
$\phi_1=\phi_3=\pi-\alpha_{0A}$=-1.90$\pi$ for different tan$\beta$. One
sees that the overlap where the $d_e$ and $d_n$ EDM constraints are
simultaneously satisfied disappears for tan$\beta>5$ for these parameters.
Note also that while $d_n$ can tolerate larger $\theta_B$, to have both EDM
constraints satisfied requires $\theta_B\stackrel{<}{\sim}0.15$ in this
example. In general, the overlap region broadens in $\Theta_1$ the closer
$\phi_1=\phi_3$ is to $-2\pi$ (i.e. real $\tilde m_1=\tilde m_3$), but the
allowed $\theta_B$ then becomes smaller (since the amoumnt of
neutralino-gluino cancelation is reduced).

\section{Models with small phases}
Both the SUGRA and D-brane models posseses a serious fine tuning problem at
$M_G$ in $\theta_{0B}$ when phases are large unless tan$\beta$ is small. For
models of this type ``naturalness" is to be defined at $M_G$, and one might
ask whether in fact models might exist where all the phases are naturally
small, e.g. $O(10^{-2})$, thus resolving the EDM problem. We present now one
such possibility.

Below the compactification sacle $M_c$, one may analyse a model in terms of
the supergravity functions of the chiral fields $\phi_\alpha$: the gauge
kinetic function $f_{ij}(\phi_{\alpha})$, the Kahler potential
$K(\phi_{\alpha},\phi_{\alpha}^{\dagger})$ and the superpotential
$W(\phi_{\alpha})$. While the origin of supersymmetry breaking remains
unknown, one may characterize it by assuming the existence of a hidden
sector where some fields, e.g. moduli or dilaton, grow VEVs of Planck mass
size:
\begin{eqnarray} x_i&=&\kappa <\Phi_i>=O(1).
\label{xi}
\end{eqnarray} Here $\kappa^{-1}=M_{Pl}=2.4\cdot 10^{18}$ GeV. We
write
$\{\Phi_{\alpha}\}=\{\Phi_i, \Phi_a\}$ where $\Phi_a$ are the physical
sector
fields and expand the Kahler potential in a power series of the physical
fields:
\begin{eqnarray} K&=&\kappa^{-2}c^{(0)}+(c_{ab}^{(2)}\Phi_a\Phi_b+{1\over M}
c_{abc}^{(3)}\Phi_a\Phi_b\Phi_c +...) +(\tilde
c_{ab}^{(2)}\Phi_a\Phi_b^{\dag}\\\nonumber
&+&{1\over M}                  
\tilde c_{abc}^{(3)}\Phi_a^{\dag}\Phi_b\Phi_c+...)
\label{kahler}
\end{eqnarray} where $M$ is a large mass. The $c^{(i)}, \tilde c^{(i)}$ are
dimensionless functions of $x_i$ and are assumed  to be $O(1)$. The first parenthesis is
holomorphic and hence can be transferred to the superpotential by a Kahler
transformation:
\begin{equation} W\rightarrow W +
\kappa^2W(c_{ab}^{(2)}\Phi_a\Phi_b+{1\over                  
M}c_{abc}^{(3)}\Phi_a\Phi_b\Phi_c+...)
\label{W}
\end{equation} The leading terms on the right arise when $W$  is
replaced by its
VEV after SUSY breaking, ($\kappa^2<W>\simeq m_{3/2}$) and one of the cubic
terms 
(e.g.
$\Phi_c$) has a GUT scale VEV arising e.g. from the GUT group breaking to
the Standard Model, $<\Phi_c>\simeq M_G$. This then  gives
 rise to the $\mu$ term, $W\rightarrow W+\mu_{ab}\Phi_a\Phi_b$,
where 
\begin{eqnarray}
\mu_{ab}&=&(c_{ab}^{(2)}+{M_G\over M}c_{abc}^{(3)})m_{3/2}.
\label{muab}
\end{eqnarray}
If one assumes now that the renormalizable terms $c^{(2)}_{ab}$ in $K$ is
real, $c^{(3)}_{abc}$ is complex but with arbitrary size phase, $\mu^0$ will
grow a phase $\theta_{0\mu}$ of size $M_G/M$ for $M>>M_G$ without any
fine tuning. For SUGRA models one expects $M=M_{\rm Pl}$ and hence
$\theta_{0\mu}\simeq O(10^{-2})$. Similar results can occur for the phases
in the soft breaking 
parameters and one might have a model where all phases are  $O(10^{-2})$
naturally suppressing the EDMs without any fine tuning. For D-brane models
 considered here one expects $M=M_{\rm str}$ and as discussed in Sec III,
 $M_{\rm str}\cong 8\times 10^{16}$ leading to phases of size
 $O(10^{-1})$. There is then a partial suppression of the EDMs leading
 perhaps to interesting predictions for the next round of EDM measurements
 for such models.
 
\section{CONCLUSIONS}
The very strong experimental constraints on the electron and neutron
electric dipole moments put additional constraints on the parameter space of
the SUSY models if the mass spectrum lies below $\sim$ 1 TeV. We have
studied this within the framework of models where the physics is determined
at a high scale (e.g. GUT or Planck scale). For SUGRA GUT models the
renormalization group equation naturally suppress $\alpha_t$, the $A_t$
phase at the electroweak scale, due to the nearness of the t-quark Landau
pole and so the phase $\alpha_{0A}$ at $M_G$ can be naturally large, even
$\pi/2$, and still lead to acceptable EDMs. The possible cancelations of
different parts of the EDM amplitudes does allow $\theta_B$, the B phase at
the electroweak scale, to be large, i.e. $O(10^{-1})$ when
$\alpha_{0A}=O(1)$, but only for low tan$\beta$ i.e. tan$\beta\le$3. The
combined conditions of the experimental EDM constraints and the requirement
of radiative breaking of $SU(2)\times U(1)$ then leads to $\theta_{0B}$ to
be $O(1)$ at the $M_G$ and very tightly determined for 
tan$\beta\stackrel{>}{\sim}3$. Thus there is a new fine tuning problem at
$M_G$ unless tan$\beta$ is small. We note that LEP data already requires
tan$\beta>2$, and the RUN II at the Tevatron will be able to probe higher
tan$\beta$ as it searches for the Higgs.

For a class of D brane models arising in Type IIB orientifolds, one can have
the gaugino mass phases obey $\phi_1=\phi_2\neq\phi_3$ allowing $\theta_B$
to become larger due to additional cancelations between gluino and
neutralino diagrams. However, the same fine tuning problem for $\theta_{0B}$
at $M_G$ arises unless tan$\beta$ is small. Further, one needs 
tan$\beta\stackrel{<}{\sim}5$ to get a significant overlap between the
allowed $d_e$ and $d_n$ regions in parameter space when the $\phi_i$ are
large.

The fact that the fine tuning problem of $\theta_{0B}$ appears to be endemic
leads one to  consider the possibility that all phases might naturally be small
at $M_G$. A simple model showing this might arise where the phases
$\phi_i=O(M_G/M)$ is discussed, where $M=M_{\rm Planck}$ for SUGRA models
and $M=M_{\rm string}$ for D-brane models.

\section{Acknowledgement}
This work was supported in part by National Science Foundation Grant No.
PHY-9722090.

\begin{figure}[htb]
\centerline{ \DESepsf(edmfig2.epsf width 8 cm) }
\smallskip
\caption {\label{fig2} Two loop Barr-Zee diagrams where A is the CP odd Higgs and
$\tilde q_i$ are the mass diagonal squark states.}
\end{figure}\begin{figure}[htb]
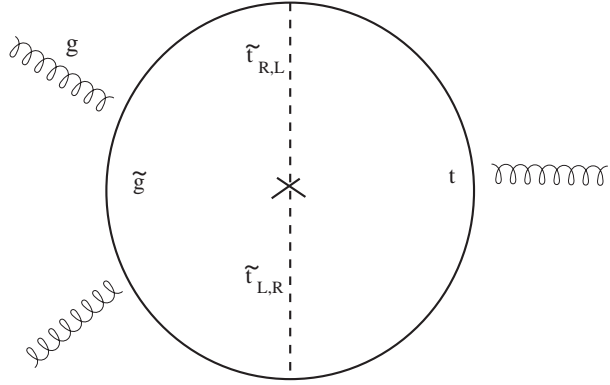

\centerline{ \DESepsf(edmfig3.epsf width 8 cm) }
\smallskip
\caption {\label{fig3} Two loop Weinberg type diagram.}
\end{figure}
\begin{figure}[htb] 
\centerline{ \DESepsf(edmfig4.epsf width 8 cm) }
\smallskip
\caption {\label{fig4} K vs $m_0$
for $d_e$ (electron EDM) for
tan$\beta=3$ (solid), 10 (dashed) and 20 (dotted) respectively for
$m_{1/2}=300$, $|A_0|= 300$ GeV,
$\alpha_{0A}={\pi\over 2}$, $\theta_B=$0.02.}
\vspace{0 cm}
\end{figure}
\begin{figure}[htb]
\centerline{ \DESepsf(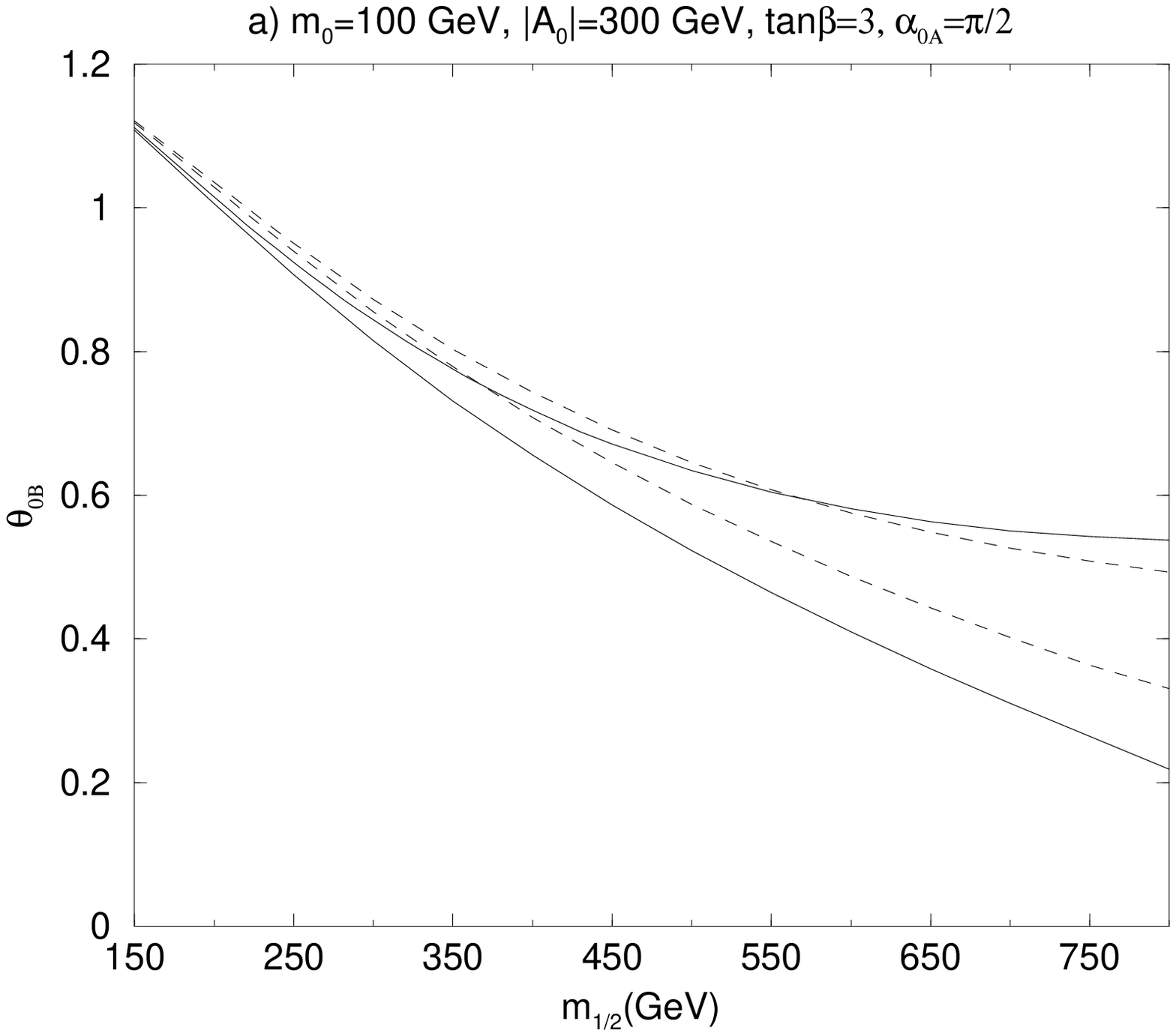 width 8 cm) } 
\centerline{ \DESepsf(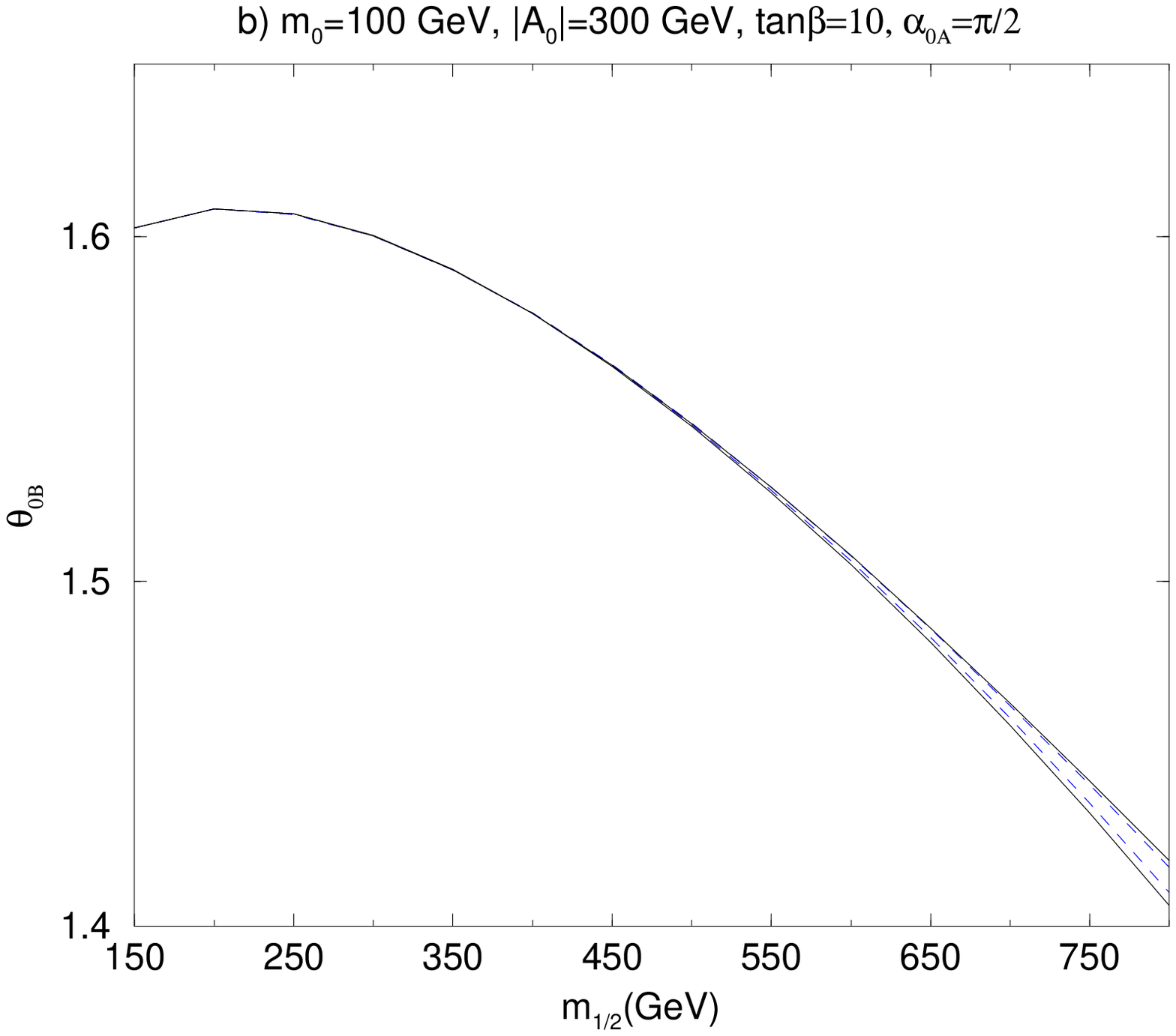 width 8 cm) }
\caption {\label{fig5}$\theta_{0B}$ vs $m_{1/2}$. Upper and lower lines are the
allowed range so that $K\le 0$. The solid lines are for  $d_n$ and the dotted
lines are for $d_e$.}
\vspace{0 cm}
\end{figure}
\begin{figure}[htb]
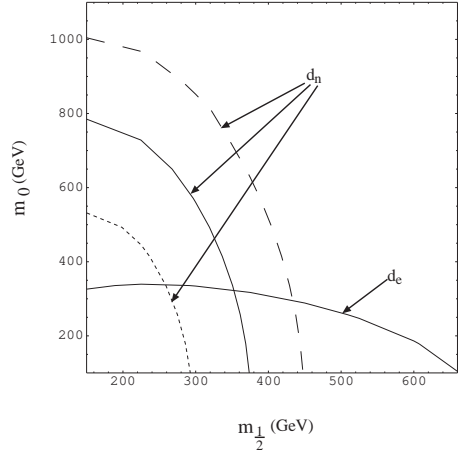
 
\centerline{ \DESepsf(edmfig9.epsf width 6 cm) }
\smallskip
\caption {\label{fig6}The K=0 contours are plotted as a function of $m_0$ and
$m_{1/2}$. The dotted, solid and dashed lines are  for $m_d$ (1 GeV)=5, 8 and 12
MeV respectively. The other input parameters are
$\alpha_{0A}={\pi\over 2}$, $|A_0|= 300$ GeV $\theta_B$=0 and tan$\beta$=3.
Excluded regions are below $d_e$ and to the left of $d_n$ curves.}
\vspace{0 cm}
\end{figure}
\begin{figure}[htb]
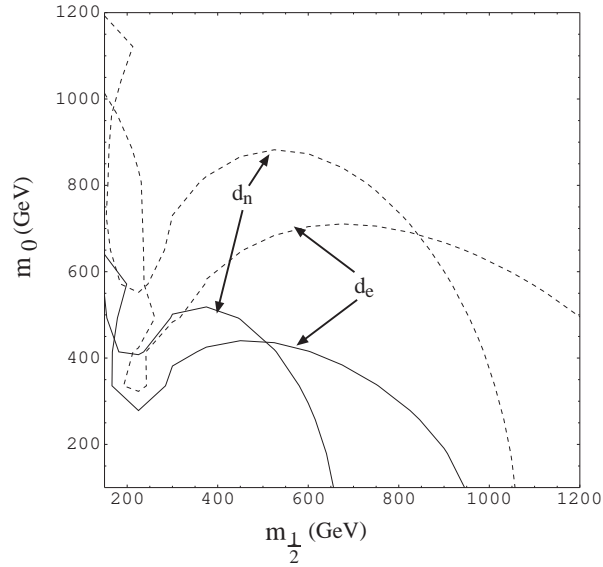
 
\centerline{ \DESepsf(edmfig10.epsf width 8 cm) }
\smallskip
\caption {\label{fig7} Allowed region in $m_0-m_{1/2}$ plane for $d_e$ and
$d_n$.
 The other input parameters are
$\alpha_{0A}={\pi\over 2}$, $|A_0|= 300$ GeV, $\theta_B$=0 and tan$\beta$=20.
 The solid lines are for K=0 and the dotted lines are for K=-0.5.}
\vspace{0 cm}
\end{figure}
\begin{figure}[htb]
\centerline{ \DESepsf(aad2edmfig5.epsf width 8 cm) }
\smallskip
\caption {\label{fig8} K vs. $\theta_B$ for $d_e$ for 
$\phi_1$=$\phi_3$=$\pi+\alpha_{0A}$=-1.25$\pi$, 
$m_{3/2}$=150 GeV, $\theta_b$=0.2, $\Theta_1=0.85$, with tan$\beta$=2(solid), 5(dashed),
10(dotted).}
\end{figure}
\begin{figure}[htb]
\centerline{ \DESepsf(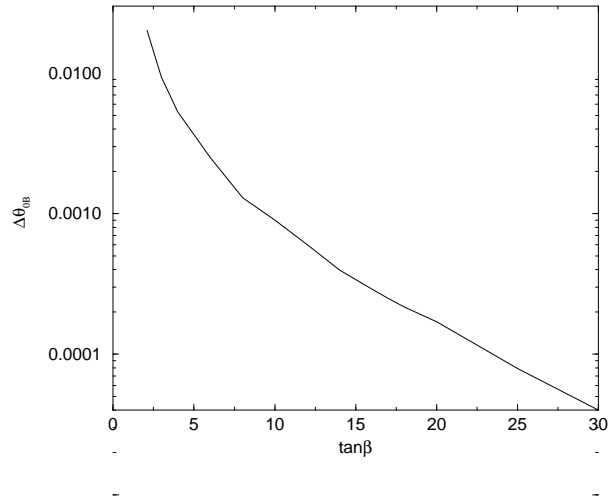 width 8 cm) }
\smallskip
\caption {\label{fig9} Values of $\Delta\theta_{0B}$ for $d_e$ satisfying the EDM
constraint as a function of tan$\beta$. Parameters are as in
Fig.8.}
\end{figure}
\begin{figure}[htb]
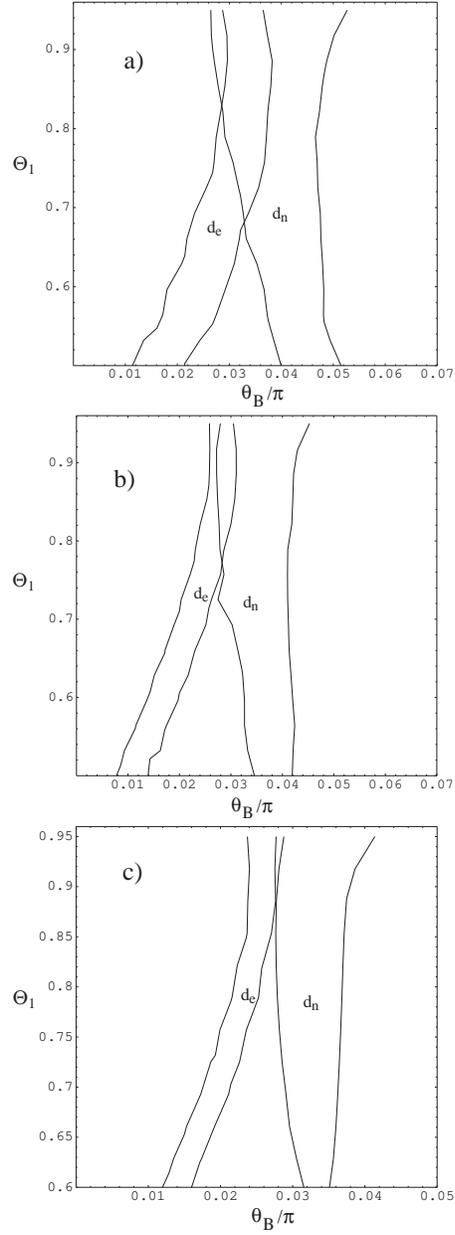

\centerline{ \DESepsf(aad2edmfig219new.epsf width 6 cm) }
\centerline{ \DESepsf(aad2edmfig319new.epsf width 6 cm) }
\centerline{ \DESepsf(aad2edmfig519new.epsf width 6 cm) }
\smallskip
\caption {\label{fig10}Allowed regions for $d_e$ and $d_n$
  for $\theta_b$=0.2, $m_{3/2}$=150 GeV and
 $\phi_1$=$\phi_3$=-1.90$\pi$, for a) tan$\beta$=2, b) tan$\beta$=3 and c)
tan$\beta$=5.}
\end{figure}

\end{document}